\documentclass[conference]{IEEEtran}
\usepackage{cite}
\usepackage{graphicx}
\usepackage{color}
\usepackage{caption}
\usepackage{subcaption}
\usepackage{epsfig}
\usepackage{epstopdf}
\usepackage{tabularx}
\usepackage{float}
\usepackage{multicol}
\usepackage{array} % for better arrays (eg matrices) in maths
\usepackage{paralist} % very flexible & customisable lists (eg. enumerate/itemize, etc.)
\usepackage{verbatim}
\usepackage[cmex10]{amsmath}
\usepackage{amssymb}
\usepackage{algorithm}
\usepackage{algorithmic}
\usepackage{multirow}
\usepackage{booktabs}
\usepackage{mdwmath}
\usepackage{mdwtab}
\usepackage{fixltx2e}
\usepackage{stfloats}
\usepackage[section]{placeins}
\normalsize
\begin{document}
\title{Secret Key Generation Based on AoA Estimation for Low SNR Conditions}

%\author{\IEEEauthorblockN{Ahmed Badawy}
%\IEEEauthorblockA{Electrical Engineering Dept.\\
%Qatar University\\
%Doha, Qatar 2713\\
%Email: badawy@qu.edu.qa}
%\and
%\IEEEauthorblockN{Tamer Khattab}
%\IEEEauthorblockA{Electrical Engineering Dept.\\
%Qatar University\\
%Doha, Qatar 2713\\
%Email: tkhattab@qu.edu.qa}
%\and
%\IEEEauthorblockN{Tarek El Fouly}
%\IEEEauthorblockA{Computer Engineering Dept.\\
%Qatar University\\
%Doha, Qatar 2713\\
%Email: tarekfouly@qu.edu.qa}
%}

\author{\IEEEauthorblockN{Ahmed Badawy\IEEEauthorrefmark{1}\IEEEauthorrefmark{2},
Tamer Khattab\IEEEauthorrefmark{2},
Tarek ElFouly\IEEEauthorrefmark{3},
Amr Mohamed\IEEEauthorrefmark{3},
Daniele Trinchero\IEEEauthorrefmark{1}\\ and
Carla-Fabiana Chiasserini \IEEEauthorrefmark{1}
}
\IEEEauthorblockA{\IEEEauthorrefmark{1}Politecnico di Torino, DET. (ahmed.badawy, daniele.trinchero, chiasserini@polito.it)}
\IEEEauthorblockA{\IEEEauthorrefmark{2}Qatar University, Electrical Engineering Dept.
(tkhattab@qu.edu.qa)}
\IEEEauthorblockA{\IEEEauthorrefmark{3}Qatar University, Computer Engineering Dept. (tarekfouly,amrm@qu.edu.qa)}
%\IEEEauthorblockA{\IEEEauthorrefmark{4}Tyrell Inc., 123 Replicant Street, Los Angeles, California 90210--4321}
}
\maketitle
\begin{abstract}
%\boldmath
In the context of physical layer security, a physical layer characteristic is used as a common source of randomness to generate the secret key. Therefore an accurate estimation of this characteristic is the core for reliable secret key generation. Estimation of almost all the existing physical layer characteristic suffer dramatically at low signal to noise (SNR) levels. In this paper, we propose a novel secret key generation algorithm that is based on the estimated angle of arrival (AoA) between the two legitimate nodes. Our algorithm has an outstanding performance at very low SNR levels. Our algorithm can exploit either the Azimuth AoA to generate the secret key or both the Azimuth and Elevation angles to generate the secret key. Exploiting a second common source of randomness adds an extra degree of freedom to the performance of our algorithm. We compare the performance of our algorithm to the algorithm that uses the most commonly used characteristics of the physical layer which are channel amplitude and phase. We show that our algorithm has a very low bit mismatch rate (BMR) at very low SNR when both channel amplitude and phase based algorithm fail to achieve an acceptable BMR.
\end{abstract}
\section{Introduction}
%The broadcast nature of the wireless communications imposes the risk of information leakage to adversarial users or unauthorized receivers. Therefore, information security between the intended users remains a challenging issue. Currently, the security relies on cryptographic techniques and protocols that lie at the upper layers of the wireless network. One main drawback of these existing techniques is the necessity of a complex key management scheme in the case of symmetric ciphers and high computational complexity in the case of asymmetric ciphers. On the other hand, physical layer security relies on the randomness of the communication channel and has a much lower computational complexity.

%One well known characteristic of the communication channel is reciprocity. When two antennas communicate by radiating the same signal through a linear and isotropic channel, the received signals by each antenna will be identical. This is mainly because of the reciprocity of the radiating and receiving antenna pattern \cite{smith04}.

Within the paradigm of physical layer secrecy, typically a physical layer specific characteristic is used as key generator to guarantee information hiding from eavesdroppers. Current physical layer security techniques are based on channel reciprocity assumption. In \cite{Azimi-Sadjadi,Li:2006,zhangg10,Mathue08, Wilson07}, channel measurements were exploited to generate the secret key. One main drawback of exploiting the channel reciprocity to generate the secret key is that the additive white Gaussian noise (AWGN) at both the receivers affects the reciprocity of the channel measurements. Also, both nodes must collect the measurement simultaneously \cite{Patwari10}.

Moreover, the techniques that exploit the channel gain, are based on the assumption that the channel gain is independent of the distance. This assumption could be valid for non-line of sight fading channel but not necessarily a valid assumption for line of sight fading channel where attenuation is a function of the propagation distance. In this case, an eavesdropper with localization or distance estimation capabilities can then estimate the channel gain and consequently recover the secret key. Others exploit the channel phase to generate the secret key as in \cite{Qian12}. For an accurate estimation of the channel phase, a high SNR is required \cite{Jana09}.

Other reciprocal (common) parameters such as received signal strength (RSS) can be used as a common source of randomness to generate the secret key \cite{Premnath13,Kitaura07, Yunchuan12}. RSS is a very common metric that requires a simple circuitry to be implemented. Nevertheless, its practical utilization as a common source of randomness is limited because its key bit generation rate is very low, particularly, for mobile scenarios \cite{Kui11_survey}.

A recent physical layer security technique that is based on the distance reciprocity to generate secret key bits is presented in \cite{Gungor11,Gungor14}. Most of the currently deployed localization technique exploit the RSS to estimate the distance between the two communicating nodes \cite{Patwari03}. Estimating the distance based on RSS requires an accurate modelling of the channel between the nodes. Moreover, it has a low estimation accuracy. In \cite {AlAlawi11}, the distance estimation error was higher than 20\%. This implies that the secret key generated based on distance will have a high bit mismatch rate (BMR), which is the ratio of the bits that do not match at the two nodes as extracted from the estimated distance. There are other techniques to perform localization which are based on the time of arrival (TOA) \cite{Zhang05,Yong02,Mailaender08,Qi06}. Although localization based on TOA has a higher accuracy than RSS based, it requires a clock synchronization between the two nodes \cite{Gezici05}. Nevertheless, their estimation error is high at low SNR ($<0$ dB) \cite{Dashti09}.

A main drawback in almost all of the existing physical layer security techniques, whether it is based on channel gain, RSS or distance,  is their poor performance at low signal to noise ratio (SNR). Estimating the channel gain at low SNR levels will result in a high error due to the effect of the AWGN. Similarly, for the RSS and consequently distance estimation based on the RSS.

To address this latter drawback, we propose a novel algorithm that exploits the AoA between the two communicating nodes.  AoA estimation techniques can accurately function even at very low SNR level. In addition to that we use the 2-D AoA (azimuth AoA and elevation AoA), which is estimated simultaneously, as a double common source of randomness. In other words, we estimate two common sources of randomness simultaneously. Exploiting a second common source of randomness adds an extra degree of freedom and increases the entropy of the generated secret key. To the best of the authors' knowledge, exploiting the AoA as a common source of randomness has not been presented in the literature before.

The rest of this paper is organized as follows: In Section II the system model is presented. The AoA estimation is then addressed in section III. Our secret key generation algorithm is presented in Section IV. We evaluate the performance of our algorithm in Section V. The paper is then concluded in section VI.

\section{System Model}
Let us assume that the two legitimate nodes, Alice and Bob, exchange a signal $s(t)$. Each of Alice's or Bob's receiver is equipped with a smart antenna system consisting of $M$ antenna elements, separated by a fixed separation $d$ and operating at frequency $f$. When using $M$ receivers, the received and sampled signal $x[n]$ in the matrix notation is:
\begin{align}
\mathbf{ X = a s + V},
\end{align}
where  $\mathbf{X}$ is of size $M \times N$ with $N$ being the total number of received samples, $\mathbf{s}$ is of size $1\times N$ as seen from each receiver, the steering vector $\mathbf{a}$ is of size $M \times 1$ and $\mathbf{V}$ is the AWGN matrix of size $M \times N$.

When using a single receiver to estimate the AoA as in our newly developed Cross Correlation Switched Beam System (XSBS) presented in \cite{badawyAoA}, the received signal reduces to:
\begin{align}
\mathbf{ x_k = a S + v},
\end{align}
where  $\mathbf{x_k}$, the received signal from the $k^{th}$ beam, is of size $1 \times N$, where $k \in [1:K]$, where $K$ is the total number of generated beams, $\mathbf{S}$ is of size $M \times N$ as seen by the $M$ elements of the antenna array and $\mathbf{v}$ is of size $1 \times N$.

Each antenna array has an array response vector also known as \textit{steering vector} $\mathbf{a}(\phi,\theta)\in \mathbb{C}^{M }$, where $\phi$ is the azimuth angle and $\theta$ is the elevation angle. For a uniform circular array (UCA), $\mathbf{a}(\phi, \theta)$, can be given by \cite{Ioannides05}:

%\begin{gather}
\begin{eqnarray}
%\begin{multlined}
\mathbf{a}(\phi, \theta)&=& [e^{\beta r \sin(\theta) \cos (\phi-\phi_1)}, e^{\beta r \sin(\theta) \cos (\phi-\phi_2)},\label{eqn_ste} \\ \nonumber
&&..., e^{\beta r \sin(\theta) \cos (\phi-\phi_M)} ],
%\end{multlined}
\end{eqnarray}
%\end{gather}

\noindent where $\beta=\frac{2\pi}{\lambda}$ is the wave number, $\lambda$ is the wavelength and $r$ is the radius of the antenna array.
\begin{align}
\phi_m=\frac{2\pi m}{M}, \hspace{0.25cm} m=1,2,..,M ,
\end{align}
and $\phi$ ranges between $[0,2\pi]$ and $\theta$ ranges between $[0,\pi]$

\begin{figure}
\begin{center}
\includegraphics[width=2.75in]{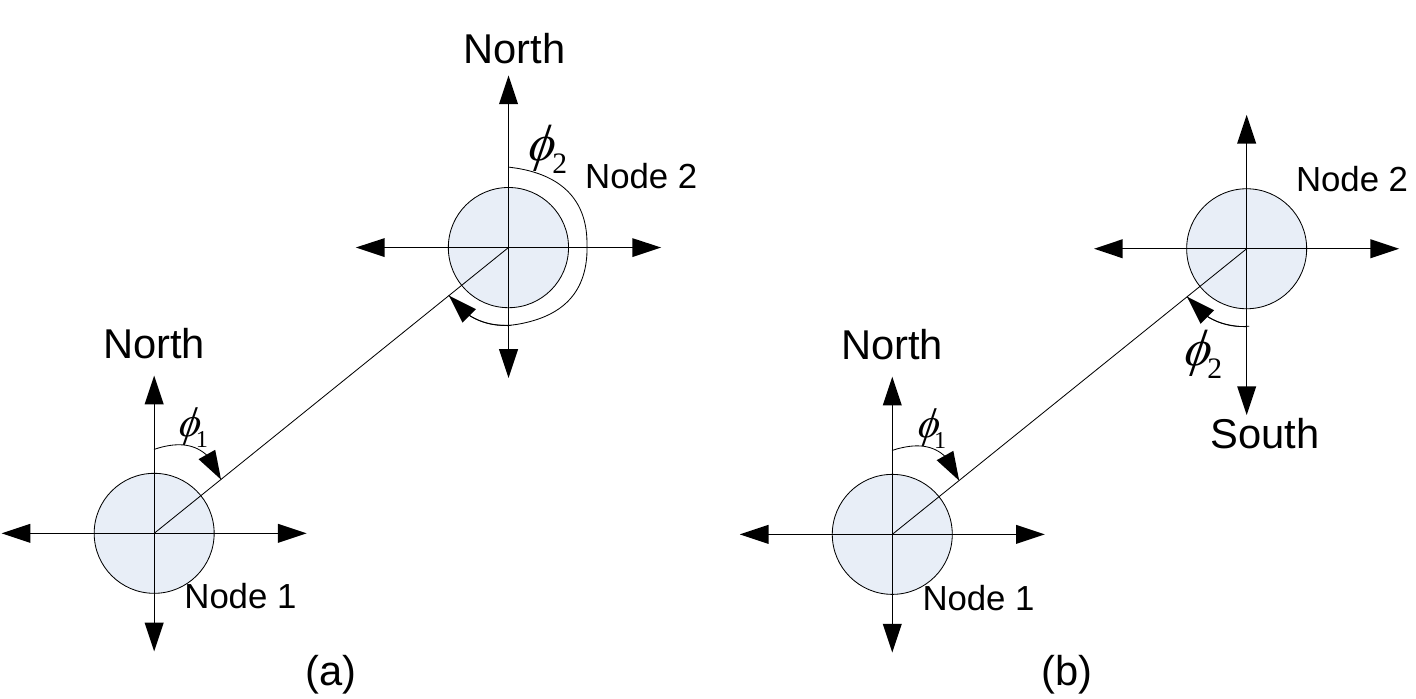}
\caption{AoA estimation reference: (a) Both have the same reference, let it be the North  and (b) Alice has the reference as the North and Bob has the reference as the South.}
\label{fig1}
\end{center}
\end{figure}

To generate a secret key based on the estimated AoA, the estimated AoA has to be common at both Alice and Bob. In other words, both Alice and Bob estimate the same AoA, whether it is 1-D (Azimuth only) only or 2-D (Azimuth and Elevation). To do so, Both Alice and Bob agree only once on a selected reference, let it be the North, along with a rotation direction, let it be Clockwise as shown in Fig. \ref {fig1} (a). In this case, the estimated AoA at Alice $\phi_1$ is:
\begin{align}
\phi_1=\phi_c,
\end{align}
where $\phi_c$ is the common AoA and the estimated AoA at Bob $\phi_2$ is:
\begin{align}
\phi_1=\phi_c + \pi
\end{align}
Therefore, Bob estimates the common AoA, simply, by subtracting $\pi$ from its estimated AoA $\phi_2$. Another approach is that Alice uses the selected reference, let it be the North and Bob uses the opposite reference which is in this case the South. The rotation direction for Both is still the same, let it be Clockwise. As shown in Fig. \ref{fig1} (b), the estimated AoAs are:
\begin{align}
\phi_1=\phi_2=\phi_c.
\end{align}
To generate a sequence of AoA and use that sequence to generate the secret key, at least one of the communication nodes, i.e., either Alice or Bob, is assumed to be mobile.
\section{AoA Estimation Techniques}
There exists many techniques to estimate the AoA; some of which are: beam switching, classical AoA techniques and subspace techniques \cite{AoA_book_Tuncer, AoA_book_Balanis, AoA_book_Chen, AoA_book_chandran, AoA_book_foutz}. Subspace based techniques perform better than classical techniques, particularly at low SNR levels. This comes on the cost that they require a higher computational complexity. The most popular AoA estimation subspace based technique is the MUltiple SIgnal Classification (MUSIC) presented in \cite{Phd1981}. For 1-D AoA estimation, the elevation angle $\theta$ is assumed to be $90$ degrees. Therefore, the steering vector for the UCA in (\ref{eqn_ste}) reduces to:
\begin{eqnarray}
%\begin{multlined}
\mathbf{a}(\phi)&=& [e^{\beta r  \cos (\phi-\phi_1)}, e^{\beta r \cos (\phi-\phi_2)},\\
&&..., e^{\beta r \cos (\phi-\phi_M)} ], \nonumber
%\end{multlined}
\end{eqnarray}
The auto-covariance matrix of the received signal, $R_{xx}$ has a dimension $M \times M$, i.e. $M$ receivers are used. $R_{xx}$ is estimated as:
\begin{align}
\mathbf{R_{xx}}=\frac{1}{N}\left(\mathbf{X} \mathbf{X^H}\right)\label{eqn10}
\end{align}
where $H$ denotes the Hermitian matrix operation. The MUSIC algorithm exploits the orthogonality of the signal and noise subspaces. After an eigenvalue decomposition (EVD) on $\mathbf{R_{xx}}$, it can be written as:
\begin{eqnarray}
\mathbf{R_{xx}}&=&\mathbf{a}(\phi) \mathbf{R_{ss}} \mathbf{a^H} (\phi)+ \sigma^2 I\\
&=& U_s \Lambda_s U_s^H +  U_v \Lambda_v U_v^H,
\end{eqnarray}

\noindent where $\mathbf{R_{ss}}$ is the autocovariance matrix of the transmitted signal, $\sigma^2$ is the noise variance, $I$ is the unitary matrix, $U_s$ and $U_v$ are the signal and noise subspaces unitary matrices and $\Lambda_s$ and $\Lambda_v$ are diagonal matrices of the eigenvalues of the signal and noise. The spatial power spectrum for the MUSIC technique is given by \cite{Phd1981,Phd2}:
\begin{align}
  P_{\text{MUSIC}}(\phi)= \frac{1}{\mathbf{a}^H(\phi) {P_v} \mathbf{a}(\phi)},
\end{align}

\noindent where $P_v = U_v U_v^H$.

Our XSBS collects an omni-directional reference signal, $\mathbf{x_o}$, using a number of antennas in the antenna array with setting the elements of the steering vector, $\mathbf{a}(\phi)$, equal to unity at selected elements (the antenna elements used as omni-directional antennas) and equal to zero in the rest. Our XSBS then starts to scan the angular region of interest and collect the signals $\mathbf{x_k}$, for $k\in[1:K]$.
The cross correlation coefficient between our omni-directional reference signal and the $k^{th}$ signal, which is our XSBS spatial power spectrum, can be given by:
\begin{align}
\mathbf{R_{ko}}=\frac{1}{N}\left(\mathbf{x_k} \mathbf{x_o^H}\right) \label{eqn11}
\end{align}

There are several ways to estimate the 2-D AoA as presented in \cite{Jun09,Hua91,Harabi07} where they use the cross correlation between the received signal from an L-shaped antenna array. In \cite{Sujuan04}, they estimate the 2-D using a UCA based on the fourth order cumulant of the the received signals. Another example in \cite{Albagory13}, they use an antenna array that consists of a vertical linear array to estimate $\theta$ using the MUSIC algorithm, they then use a circular antenna array with $\theta$ fed to the MUSIC algorithm again to estimate $\phi$.

\begin{figure}
\centering
\includegraphics[width=3in]{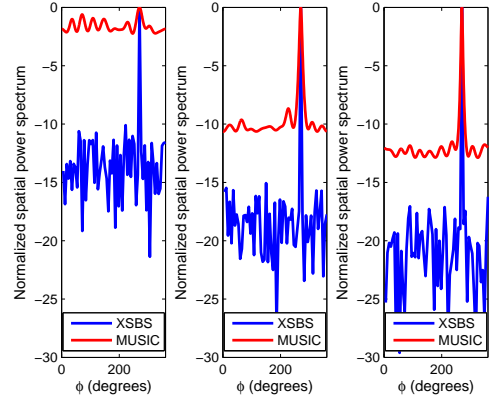}
\caption{Spatial power spectrum of MUSIC vs. XSBS for $\phi=270$ degrees at SNR = -15 dB for $N=100$ samples (left), $N= 1000$ samples (middle) and $N=2000$ samples (right). }
\label{fig_3}
\end{figure}

Figure \ref{fig_3} shows the simulation results for both the MUSIC algorithm for $M=16$, and for XSBS for $M=17$, with five antenna used as omni-directional antennas to collect $\mathbf{x_o}$ with a separation between each two antennas of $2*d$ to such that the correlation between the signals received for the selected antenna elements is minimized. The simulation results are for $\phi = 270$ degrees using a UCA and $N=100$ samples (left), $N= 1000$ samples (middle) and $N=2000$ samples (right). The simulation is at SNR = - 15 dB. One can see that both algorithms have a remarkable performance at SNR levels as low as -15 dB. The MUSIC algorithm is achieving a peak to floor ratio (PFR) of  3, 10 and 13 for $N=100$, $N=1000$ and $N=2000$, respectively. On the other hand the PFR for the XSBS 15, 19, and 23 for $N=100$, $N=1000$ and $N=2000$, respectively.
Increasing the number of samples enhances the performance of both algorithms. For an adequate number of collected samples $N=1000$, both algorithms will have a decent performance even at very low SNR levels.
%\begin{figure}
%\centering
%\includegraphics[width=2.75in]{MUSIC_vs_XSBS_M_10_different_samples_phi_270.eps}
%\caption{Spatial power spectrum of MUSIC vs. XSBS for $\phi=270$ degrees at SNR = -10 dB for $N=100$ samples (left), $N= 1000$ samples (middle) and $N=2000$ samples (right). }
%\label{fig_2}
%\end{figure}
\section{Secret Key Generation Algorithm}
Both Alice and Bob start exchanging signals to estimate the AoA and consequently generate the secret key. The steps to generate the secret key based on the AoA are:
\subsection{Initialization}
Both Alice and Bob agree on the reference as well as the rotation direction, from which the AoA is estimated. This step is performed only once at the beginning of communication between them. It is not required to be applied each time Alice and Bob communicate.
\subsection{AoA Estimation}
Both Alice and Bob estimate their AoA and based on the selected reference, they estimate the common source of randomness, i.e., $\phi_c$ for 1-D or $\phi_c$ and $\theta_c$ for 2-D. The algorithm applied at either Alice or Bob does not necessarily be the same. One can use the MUSIC if it can afford both the computational and hardware complexity. The other can use the XSBS if, for example, it is a portable device and can not afford both computational and hardware complexities. Both Algorithms as we showed earlier can operate in low SNR levels. Other techniques could be used as well after studying their performance at low SNRs to make sure that the generated secret key will have a low BMR.
\subsection{Quantization}
Now that we have the common sources of randomness $\phi_c$, the third step of our algorithm is to convert it into a bit stream suitable for the secret key generation. The conventional secret key length is between 128 and 512 bits \cite{Mathue08}. We use the most popular technique for quantization which is the uniform quantization \cite{LiBook07}:
\begin{align}
z=Q(y) \hspace{1in} y\in{(p_i, p_{i+1})}
\end{align}
where $p$ is the interval and $y$ is the input, which in this case is estimated AoA. In the uniform quantization, the spaces along the x-axis, i.e., time, is uniformly distributed. Similarly for the spaces in the y-axis, i.e., the estimated AoA. We use $n_{quan}$ bits and therefore $2^{n_{quan}}$ levels to quantize our common sources of randomness and then convert the quantized decimal values into bits.
\subsection{Encoding}
Although uniform quantization is easy to implement, increasing the quantization bit number, dramatically degrades the performance of the algorithm since the bit mismatch rate between the two communicating nodes increases. In \cite{zhangg10}, an encoding algorithm is proposed to tackle this problem where each uniformly quantized value is encoded with multiple values. We encode our most significant bit with $n_{encod}$ bits.
\subsection{Combining the Two Bit Streams}
Now that we have measured, quantized and encoded our two common sources of randomness, which are the elevation AoA and the Azimuth AoA, we have two bit streams containing these data. To combine these two bit streams, any logical operation such as AND, OR or concatenation can be applied on the two bit streams to generate a single bit stream containing both Azimuth and Elevation angles information. We choose to use concatenation operation with the two bit streams as the inputs to generate the single bit stream. Before we concatenate, we drop the least significant $n_{quan}-n_{combn}$ bits from each single bit stream, where $n_{comb}$ is the number of bits selected from each bit stream. \textit{It is worth noting that we chose a simple bit operation to be applied on the bit streams for the sake of simplification}. One can apply a more complicated operation at the bit streams such as bit masking or combinations of series and parallel logical gates.

\emph{Up to this step, the key is generated and ready to be used to encrypt the transmitted data. The following steps are optional and preferred to be used at very low SNR levels (below -20 dB) where the generated key will have a considerable BMR.}
\subsection{Information Reconciliation}
The generated bit streams at Alice and Bob might have some discrepancy, particularly at very low SNR levels. This is due to several reasons such as interference, noise and hardware limitations. We adopt the reconciliation protocol presented in \cite{Brassard94} to minimize the discrepancy. Both Alice and Bob first permute their bit streams in the same way. Then they divide the permuted bit stream into small blocks. Alice then sends permutations and parities of each block to Bob. Bob then compares the received parity information with the ones he already processed. In case of a parity mismatch, Bob changes his bits in this block to match the received ones.
\subsection{Privacy Amplification}
Although information reconciliation protocol leaks minimum information, the eavesdropper can still use this leaked information to guess the rest of the secret key. Privacy amplification solves this issue by reducing the length of the outputted bit stream. The generated bit stream is shorter in length but higher in entropy. To do so, both Alice and Bob apply a universal hash function selected randomly from a set of hash functions known by both Alice and Bob. Alice sends the number of the selected hash function to Bob so that Bob can use the same hash function. Our algorithm is summarized below.
\begin{algorithm}
\begin{algorithmic}
\STATE \textbf{Step 0: Initialization}
\STATE Alice and Bob agree on the reference and the rotation direction from which they estimate the AoA.
\STATE \textbf{Step 1: AoA Estimation}
\STATE Alice and Bob estimate the common source(s) of randomness, $\phi_c$, or $\phi_c$ and $\theta_c$, each using its implemented technique.
\STATE \textbf{Step 2: Uniform Quantization}
\STATE Alice and Bob quantize the $\phi_c$ or $\phi_c$ and $\theta_c$ using $n_{quan}$ bits to convert the decimal values into bits.
\STATE \textbf{Step 3: Encoding}
\STATE Alice and Bob encode each uniformly quantized value with multiple values $n_{encod}$.
\STATE \textbf{Step 4: Combining the Two Bit Streams}
\STATE Alice and Bob apply concatenate the two bit streams.
\STATE \textbf{Step 5: Information Reconciliation} (Optional for very low SNR)
\STATE Alice and Bob permute the bit stream and divide them into small blocks.
\STATE Alice sends the permutation and parities to Bob.
\STATE Bob compares the received parity information with his.
\STATE  In case of mismatch, Bob corrects his bits accordingly.
\STATE \textbf{Step 6: Privacy Amplification} (Optional for very low SNR)
\STATE Alice sends the number of the hash function to Bob.
\STATE Alice and Bob apply the hash function to the bit stream.
\end{algorithmic}
\caption{Secret Key Generation algorithm}
\label{alg1}
\end{algorithm}
\section{Performance Evaluation}
To show that the secret key generated based on the estimated AoA will have a low BMR at low SNR levels, we first plot the root mean squared error (RMSE) of the estimated AoA for the two algorithms; the MUSIC and the XSBS. The RMSE is defined as:
\begin{align}
RMSE = \sqrt{E\left((\hat{\phi_c}-\phi_c)^2\right)} \label{eqn100}
\end{align}
where $E[.]$ denotes the mean operation and $\hat{\phi_c}$ is the actual estimated AoA of the true AoA, $\phi_c$.

\begin{figure}
\centering
\includegraphics[width=2.5in]{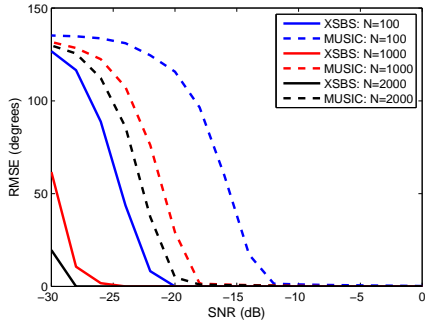}
\caption{RMSE vs. SNR for the MUSIC algorithm and for the XSBS. }
\label{fig_10}
\end{figure}

Fig. \ref{fig_10} presents the RMSE for both the MUSIC as well as the XSBS versus SNR for different number of samples for the Azimuth angle.
%Fig. \ref{fig_11} presents the RMSE for both the MUSIC as well as the XSBS versus number of samples for different SNR values.
%Both figure \ref{fig_10} and \ref{fig_11} are for the elevation angle $\phi$.
The true Azimuth angle is $\phi_c=270$ degrees and the RMSE is estimated according to Eq. (\ref{eqn100}). Table \ref{Table1} summarizes the RMSE values for both the MUSIC and the XSBS for different number of samples at different SNR values for the Azimuth angle .
%The RMSE for the elevation angle for the two algorithm is presented in Fig. \ref{fig_12}.
Table \ref{Table2} summarizes the RMSE values for both the MUSIC and the XSBS at different SNR values for the Elevation angle for $N=1000$ samples. The true Elevation angle is $\theta_c=90$ degrees and the RMSE is estimated according to Eq. (\ref{eqn100}).

From Tables \ref{Table1} and \ref{Table2}, one can see that both the MUSIC and the XSBS have a low RMSE at low SNR levels.
%Both algorithms have negligible RMSE at SNR = -10 dB when using a low number of samples, $N= 100$.
As the SNR decreases, more samples are required to achieve a very low RMSE. The XSBS outperforms the MUSIC algorithms, particularly at very low SNR levels. One can see that when using an adequate number of samples, the RMSE of both algorithm will be very low. Consequently, the secret key generated using the estimated AoA as the seed will have a low BMR.
%\begin{figure}
%\centering
%\includegraphics[width=2.5in]{RMSE_XSBS_MUSIC_vs_Samples_phi_270.eps}
%\caption{RMSE vs. samples for the MUSIC algorithm and for the XSBS. }
%\label{fig_11}
%\end{figure}
%\begin{figure}
%\centering
%\includegraphics[width=2.5in]{RMSE_XSBS_Elevation.eps}
%\caption{RMSE vs. SNR for the MUSIC algorithm and for the XSBS for the Elevation angle. }
%\label{fig_12}
%\end{figure}

%From Figures \ref{fig_10} to \ref{fig_12}, one can see that both the MUSIC and the XSBS have a low RMSE at low SNR levels. Both algorithms have negligible RMSE at SNR = -10 dB when using a low number of samples, $N= 100$. As the SNR decreases, more samples are required to achieve a very low RMSE. The MUSIC algorithm requires around 400 samples at SNR = -15 dB to achieve a negligible RMSE, while the XSBS requires 100 samples. The XSBS outperforms the MUSIC algorithms, particularly at very low SNR levels. As the number of samples increases, the performance of the two algorithms improves.

\begin{figure}
\centering
\includegraphics[width=3in]{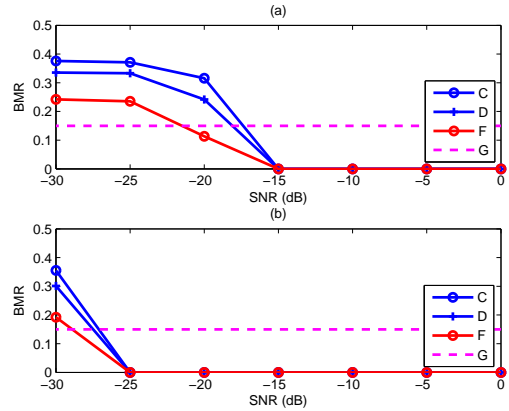}
\caption{BMR for (a) MUSIC and (b) XSBS vs. SNR for Azimuth angle, Elevation angle and both angles combined. }
\label{fig_13}
\end{figure}

We use the estimated RMSE to generate random Azimuth and Elevation angles and use them as the seed to generate the secret key. We compare the BMR of the generated keys based on AoA with the BMR of the most commonly used physical layer characteristics which are the channel gain and phase. The simulation parameters for the subsequent Figures \ref{fig_13} to \ref{fig_16} are summarized in Table \ref{Table3}. Also, the Legends for the curves within the same figures are identified in Table \ref{Table4}. We first use a single characteristic, i.e., amplitude only, phase only, Azimuth angle only and Elevation angle only. We then combine the channel amplitude and phase and combine the Azimuth and Elevation angles to generate the secret key. It's worth noting that the acceptable BMR threshold as presented in \cite{Gungor14} is 0.15 to achieve a reliability condition.

\begin{table}
\caption{RMSE for MUSIC vs. XSBS for the Azimuth angle.} \label{Table1}
\begin{tabular}{|c|c|c|c|c|c|c|}
  \hline
 % -- & -- & -- & -- & -- & -- & -- \\
  \multirow{3}{*}{SNR (dB)} & \multicolumn{6}{c|}{RMSE (degrees)}\\\cline{2-7}
  &\multicolumn{2}{c|}{N= 100} & \multicolumn{2}{c|}{N= 1000} & \multicolumn{2}{c|}{N= 2000}\\\cline{2-7}
   &MUSIC & XSBS & MUSIC & XSBS & MUSIC & XSBS \\\hline \hline
  -10 & 0 & 0 & 0 & 0 & 0 & 0 \\\hline
  -15 & 39 & 0 & 0 & 0 & 0 & 0 \\\hline
  -20 & 115 & 0 & 29 & 0 & 5 & 0 \\\hline
  -25 & 132 & 66 & 114 & 0 & 98 & 0 \\\hline
  -30 & 135 & 126 & 131 & 61 & 129 & 20 \\
  \hline
\end{tabular}
\end{table}
For a fair comparison between the different common sources of randomness, we first scale the sequence of information collected to the same scaling level such that all common sources of randomness used below, i.e., channel amplitude, channel phase, Azimuth angle and Elevation angle fluctuate within the same levels.
\begin{table}
\caption{RMSE for MUSIC vs. XSBS for the Elevation angle for $N=1000$ Samples.} \label{Table2}
\centering
\begin{tabular}{|c|c|c|}
  \hline
 % -- & -- & -- & -- & -- & -- & -- \\
  \multirow{2}{*}{SNR (dB)} & \multicolumn{2}{c|}{RMSE (degrees)} \\\cline{2-3}
   &MUSIC & XSBS   \\\hline \hline
  -10 & 0 & 0  \\\hline
  -15 & 0 & 0  \\\hline
  -20 & 8 & 0  \\\hline
  -25 & 34 & 0  \\\hline
  -30 & 37 & 20 \\
  \hline
\end{tabular}
\end{table}
\begin{table}
\caption{Simulation parameters for the subsequent figures} \label{Table3}
\begin{tabular}{|c|c|c|c|c|c|}
  \hline
 % -- & -- & -- & -- & -- & -- & -- \\

  Figure & Algorithm & Samples & Quan. Bits & Enc. Bits & Comb. Bits  \\\hline
  Fig. \ref{fig_13}  & Both & 1000 &7 & 2 & 2  \\\hline
  Fig. \ref{fig_14}& MUSIC &  1000&6:9 & 2 & 5   \\\hline
  Fig. \ref{fig_15}& MUSIC & 1000& 7 & 1:4 & 5  \\\hline
  Fig. \ref{fig_16}& MUSIC & 1000& 7 & 2 & 3:6   \\
  \hline
\end{tabular}
\end{table}
\begin{table}
\caption{Legend} \label{Table4}
\centering
\begin{tabularx} {0.48\textwidth}{|X|X|X|X|X|X|X|}
  \hline
 % -- & -- & -- & -- & -- & -- & -- \\

  A & B & C & D & E & F & G   \\\hline \hline
  Chan. amp. & Chan. phase & Az. angle. & Elev. angle & Comb. amp. \& ph & Comb. Az. \& Elev & Thresh.\\
  \hline
\end{tabularx}
\end{table}
\subsection{MUSIC vs. XSBS}
In Fig. \ref{fig_13} we compare the performance of the MUSIC algorithm versus the XSBS in generating the secret key. It can be seen that the algorithm based on XSBS outperforms the MUSIC based algorithm, which was expected since the RMSE for the XSBS is lower than that for the MUSIC. The MUSIC based algorithm can operate within the acceptable range up to - 17 dB, while the XSBS based can operate up to -27 dB.
\subsection{Effect of number of quantization bits}
The first observation aside from the effect of any parameter whether it is the number of quantization bits or the encoding bits, which can be seen from the subsequent Figures, that our AoA based algorithm significantly outperforms both the channel amplitude and phase based ones. It is shown that the our algorithm has an operating range below the acceptable threshold which varies according to the testing parameters. Unlike the channel amplitude and phase based algorithm that fail to have an operating range at that low SNR level by achieving a BMR much higher than the acceptable threshold. Also, it is worth noting that the upper bound on the BMR is 0.5 which is equivalent to random guessing. In other words, the highest, i.e., the worst BMR is 0.5.

\begin{figure}
\centering
\includegraphics[width=3.5in]{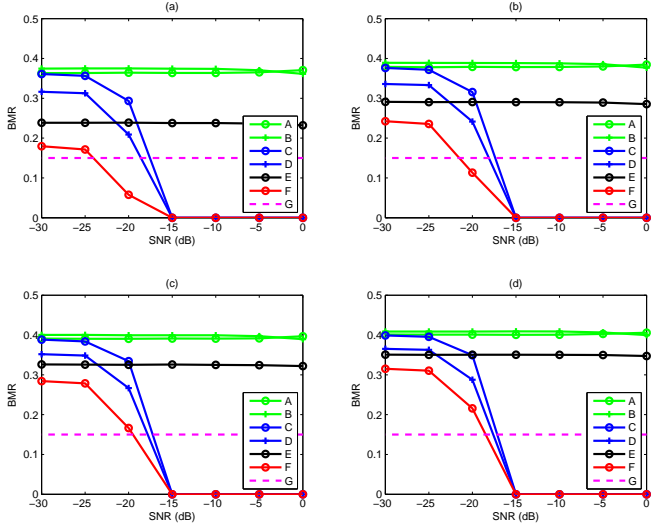}
\caption{BMR for the AoA based algorithm vs. channel based for (a) $n_{quan}=6$ and (b) $n_{quan}=7$ (c) $n_{quan}=8$ (d) $n_{quan}=9$ .}
\label{fig_14}
\end{figure}

It is shown from Fig. \ref{fig_14} that as the number of quantization bits increases, the performance of our algorithm deteriorates. This is expected since as the number of quantization bits increases, more levels are added. Therefore a smaller mismatch or error between the estimated AoAs will lead to more mismatched bit. The acceptable range using $n_{quan}=7$ is as low as -16 dB using the Azimuth angle, -17 dB using the Elevation angle and -22 using the combination of both of them.
\subsection{Effect of number of Encoding bits}
\begin{figure}
\centering
\includegraphics[width=3.5in]{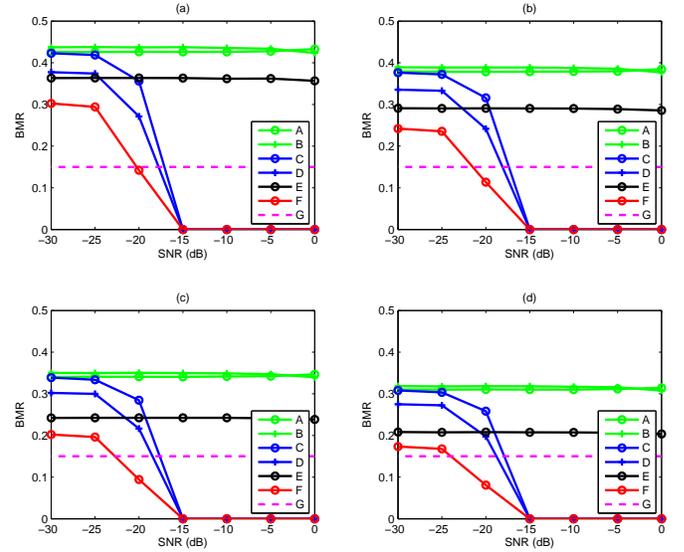}
\caption{BMR for the AoA based algorithm vs. channel based for (a) $n_{encod}=1$ and (b) $n_{encod}=2$ (c) $n_{encod}=3$ (d) $n_{encod}=4$.}
\label{fig_15}
\end{figure}

It is shown from Fig. \ref{fig_15} that as the number of encoding bits increases, the performance of our algorithm improves. As the number of encoding bits increases, more matched bits are added to soothe the effect of quantization. The acceptable range using $n_{encod}=2$ is as low as -16 dB using the Azimuth angle, -17 dB using the Elevation angle and -22 using the combination of both of them.
\subsection{Effect of number of Combining bits}
\begin{figure}
\centering
\includegraphics[width=3.5in]{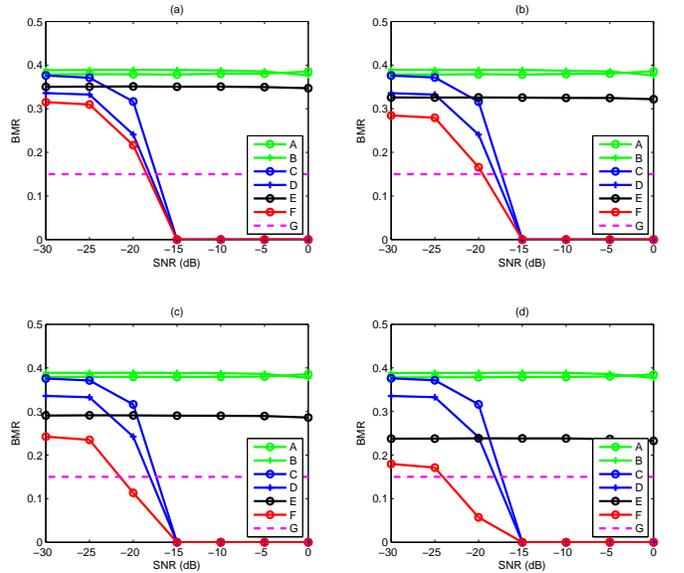}
\caption{BMR for the AoA based algorithm vs. channel based for (a) $n_{comb}=3$ and (b) $n_{comb}=4$ (c) $n_{comb}=5$ (d) $n_{comb}=6$.}
\label{fig_16}
\end{figure}

It is shown from Fig. \ref{fig_16} that as the number of combining bits increases, the performance of our algorithm improves. In addition to that, the higher the number of combining bits the longer the generated key which is the main advantage of the concatenation process. The acceptable range using $n_{comb}=5$ is as low as -16 dB using the Azimuth angle, -17 dB using the Elevation angle and -22 using the combination of both of them.
\section{Conclusion}
In this paper, we proposed a novel secret key generation algorithm that is based on the estimated AoA between the two legitimate nodes. We first showed that the RMSE for the estimated AoA between Alice and Bob is very low at very low SNR levels. We used both the 1-D AoA information and the 2-D AoA information. Exploiting a second common source of randomness adds an extra degree of freedom to the algorithm since one can use either a single common source or combine both of them in a way that minimizes the BMR. We compared the performance of our algorithm to the most widely used; the channel gain based algorithm. We showed that our algorithm has significantly outperformed the channel gain based algorithm at low SNR levels. We also studied the effect of number of quantization bits, number of encoding bit and number of combining bits on the performance of our algorithm.

\section*{Acknowledgment}
This research was made possible by NPRP 5-559-2-227 grant from the Qatar National Research Fund (a member of The Qatar Foundation). The statements made herein are solely the responsibility of the authors.
\bibliographystyle{IEEEtran}
\bibliography{references_secret_key_AOA}

\begin{thebibliography}{10}
\providecommand{\url}[1]{#1}
\csname url@samestyle\endcsname
\providecommand{\newblock}{\relax}
\providecommand{\bibinfo}[2]{#2}
\providecommand{\BIBentrySTDinterwordspacing}{\spaceskip=0pt\relax}
\providecommand{\BIBentryALTinterwordstretchfactor}{4}
\providecommand{\BIBentryALTinterwordspacing}{\spaceskip=\fontdimen2\font plus
\BIBentryALTinterwordstretchfactor\fontdimen3\font minus
  \fontdimen4\font\relax}
\providecommand{\BIBforeignlanguage}[2]{{%
\expandafter\ifx\csname l@#1\endcsname\relax
\typeout{** WARNING: IEEEtran.bst: No hyphenation pattern has been}%
\typeout{** loaded for the language `#1'. Using the pattern for}%
\typeout{** the default language instead.}%
\else
\language=\csname l@#1\endcsname
\fi
#2}}
\providecommand{\BIBdecl}{\relax}
\BIBdecl

\bibitem{Azimi-Sadjadi}
B.~Azimi-Sadjadi, A.~Kiayias, A.~Mercado, and B.~Yener, ``Robust key generation
  from signal envelopes in wireless networks,'' in \emph{Proceedings of the
  14th ACM Conference on Computer and Communications Security}, ser. CCS '07,
  2007, pp. 401--410.

\bibitem{Li:2006}
Z.~Li, W.~Xu, R.~Miller, and W.~Trappe, ``Securing wireless systems via lower
  layer enforcements,'' in \emph{Proceedings of the 5th ACM Workshop on
  Wireless Security}, ser. WiSe '06, 2006, pp. 33--42.

\bibitem{zhangg10}
J.~Zhang, S.~Kasera, and N.~Patwari, ``Mobility assisted secret key generation
  using wireless link signatures,'' in \emph{INFOCOM, 2010 Proceedings IEEE},
  2010, pp. 1--5.

\bibitem{Mathue08}
S.~Mathur, W.~Trappe, N.~Mandayam, C.~Ye, and A.~Reznik, ``Radio-telepathy:
  Extracting a secret key from an unauthenticated wireless channel,'' in
  \emph{Proceedings of the 14th ACM International Conference on Mobile
  Computing and Networking}, ser. MobiCom '08, 2008, pp. 128--139.

\bibitem{Wilson07}
R.~Wilson, D.~Tse, and R.~Scholtz, ``Channel identification: Secret sharing
  using reciprocity in ultrawideband channels,'' \emph{Information Forensics
  and Security, IEEE Transactions on}, vol.~2, no.~3, pp. 364--375, 2007.

\bibitem{Patwari10}
N.~Patwari, J.~Croft, S.~Jana, and S.~Kasera, ``High-rate uncorrelated bit
  extraction for shared secret key generation from channel measurements,''
  \emph{Mobile Computing, IEEE Transactions on}, vol.~9, no.~1, pp. 17--30,
  2010.

\bibitem{Qian12}
Q.~Wang, K.~Xu, and K.~Ren, ``Cooperative secret key generation from phase
  estimation in narrowband fading channels,'' \emph{Selected Areas in
  Communications, IEEE Journal on}, vol.~30, no.~9, pp. 1666--1674, October
  2012.

\bibitem{Jana09}
S.~Jana, S.~N. Premnath, M.~Clark, S.~K. Kasera, N.~Patwari, and S.~V.
  Krishnamurthy, ``On the effectiveness of secret key extraction from wireless
  signal strength in real environments,'' in \emph{Proceedings of the 15th
  Annual International Conference on Mobile Computing and Networking}, ser.
  MobiCom '09, 2009, pp. 321--332.

\bibitem{Premnath13}
S.~N. Premnath, S.~Jana, J.~Croft, P.~L. Gowda, M.~Clark, S.~K. Kasera,
  N.~Patwari, and S.~V. Krishnamurthy, ``Secret key extraction from wireless
  signal strength in real environments,'' \emph{Mobile Computing, IEEE
  Transactions on}, vol.~12, no.~5, pp. 917--930, 2013.

\bibitem{Kitaura07}
A.~Kitaura, H.~Iwai, and H.~Sasaoka, ``A scheme of secret key agreement based
  on received signal strength variation by antenna switching in land mobile
  radio,'' in \emph{Advanced Communication Technology, The 9th International
  Conference on}, vol.~3, Feb 2007, pp. 1763--1767.

\bibitem{Yunchuan12}
Y.~Wei, C.~Zhu, and J.~Ni, ``Group secret key generation algorithm from
  wireless signal strength,'' in \emph{Internet Computing for Science and
  Engineering (ICICSE), 2012 Sixth International Conference on}, April 2012,
  pp. 239--245.

\bibitem{Kui11_survey}
K.~Ren, H.~Su, and Q.~Wang, ``Secret key generation exploiting channel
  characteristics in wireless communications,'' \emph{Wireless Communications,
  IEEE}, vol.~18, no.~4, pp. 6--12, August 2011.

\bibitem{Gungor11}
O.~Gungor, F.~Chen, and C.~Koksal, ``Secret key generation from mobility,'' in
  \emph{GLOBECOM Workshops (GC Wkshps), 2011 IEEE}, 2011, pp. 874--878.

\bibitem{Gungor14}
O.~G{\"u}ng{\"o}r, F.~Chen, and C.~E. Koksal, ``Secret key generation from
  mobility,'' \emph{CoRR}, vol. abs/1112.2793, 2011.

\bibitem{Patwari03}
N.~Patwari and A.~O. Hero, III, ``Using proximity and quantized rss for sensor
  localization in wireless networks,'' in \emph{Proceedings of the 2Nd ACM
  International Conference on Wireless Sensor Networks and Applications}, ser.
  WSNA '03, 2003, pp. 20--29.

\bibitem{AlAlawi11}
R.~Al~Alawi, ``Rssi based location estimation in wireless sensors networks,''
  in \emph{Networks (ICON), 2011 17th IEEE International Conference on}, 2011,
  pp. 118--122.

\bibitem{Zhang05}
Z.~Zhang, C.~Law, and Y.~Guan, ``Ba-poc-based ranging method with multipath
  mitigation,'' \emph{Antennas and Wireless Propagation Letters, IEEE}, vol.~4,
  no.~1, pp. 492--495, 2005.

\bibitem{Yong02}
J.-Y. Lee and R.~Scholtz, ``Ranging in a dense multipath environment using an
  uwb radio link,'' \emph{Selected Areas in Communications, IEEE Journal on},
  vol.~20, no.~9, pp. 1677--1683, 2002.

\bibitem{Mailaender08}
L.~Mailaender, ``On the geolocation bounds for round-trip time-of-arrival and
  all non-line-of-sight channels,'' \emph{EURASIP Journal on Advances in Signal
  Processing}, vol. 2008, no.~1, p. 584670, 2008.

\bibitem{Qi06}
Y.~Qi, H.~Kobayashi, and H.~Suda, ``Analysis of wireless geolocation in a
  non-line-of-sight environment,'' \emph{Wireless Communications, IEEE
  Transactions on}, vol.~5, no.~3, pp. 672--681, 2006.

\bibitem{Gezici05}
S.~Gezici, Z.~Tian, G.~Giannakis, H.~Kobayashi, A.~Molisch, H.~Poor, and
  Z.~Sahinoglu, ``Localization via ultra-wideband radios: a look at positioning
  aspects for future sensor networks,'' \emph{Signal Processing Magazine,
  IEEE}, vol.~22, no.~4, pp. 70--84, 2005.

\bibitem{Dashti09}
M.~Dashti, M.~Ghoraishi, and J.-i. Takada, ``Optimum threshold for ranging
  based on toa estimation error analysis,'' in \emph{Personal, Indoor and
  Mobile Radio Communications, 2009 IEEE 20th International Symposium on}, Sept
  2009, pp. 988--992.

\bibitem{badawyAoA}
\BIBentryALTinterwordspacing
D.~T. T.~E. Ahmed~Badawy, Tamer~Khattab and A.~Mohamed, ``A simple aoa
  estimation scheme,'' \emph{CoRR}, vol. abs/1409.5744, 2014. [Online].
  Available: \url{http://arxiv.org/abs/1409.5744}
\BIBentrySTDinterwordspacing

\bibitem{Ioannides05}
P.~Ioannides and C.~Balanis, ``Uniform circular arrays for smart antennas,''
  \emph{Antennas and Propagation Magazine, IEEE}, vol.~47, no.~4, pp. 192--206,
  Aug 2005.

\bibitem{AoA_book_Tuncer}
T.~E. Tuncer and B.~Friedlander, \emph{Classical and Modern
  Direction-of-Arrival Estimation}.\hskip 1em plus 0.5em minus 0.4em\relax
  Academic Press, 2009.

\bibitem{AoA_book_Balanis}
C.~A. Balanis and P.~I. Ioannides, ``Introduction to smart antennas,''
  \emph{Synthesis Lectures on Antennas}, vol.~2, no.~1, pp. 1--175, 2007.

\bibitem{AoA_book_Chen}
Z.~Chen, G.~Gokeda, and Y.~Yu, \emph{Introduction to Direction-of-Arrival
  Estimation}.\hskip 1em plus 0.5em minus 0.4em\relax Artech House, 2010.

\bibitem{AoA_book_chandran}
S.~Chandran, \emph{Advances in Direction-of-Arrival Estimation}.\hskip 1em plus
  0.5em minus 0.4em\relax Artech House, 2006.

\bibitem{AoA_book_foutz}
J.~Foutz, A.~Spanias, and M.~K. Banavar, \emph{Narrowband Direction of Arrival
  Estimation for Antenna Arrays}.\hskip 1em plus 0.5em minus 0.4em\relax Morgan
  \& Claypool Publishers, 2006.

\bibitem{Phd1981}
R.~O. Schmidt, ``{A signal subspace approach to multiple emitter location and
  spectral estimation},'' Ph.D. dissertation, Stanford University.

\bibitem{Phd2}
A.~Khallaayoun, ``{A High Resolution Direction of Arrival Estimation Analysis
  and Implementation in a Smart Antenna System},'' Ph.D. dissertation, Montana
  State University.

\bibitem{Jun09}
N.-J. Li, J.-F. Gu, and P.~Wei, ``Simple and efficient cross-correlation method
  for estimating 2-d direction of arrival,'' in \emph{Wireless Communications,
  Networking and Mobile Computing, 2009. WiCom '09. 5th International
  Conference on}, Sept 2009, pp. 1--4.

\bibitem{Hua91}
Y.~Hua, T.~Sarkar, and D.~Weiner, ``An l-shaped array for estimating 2-d
  directions of wave arrival,'' \emph{Antennas and Propagation, IEEE
  Transactions on}, vol.~39, no.~2, pp. 143--146, Feb 1991.

\bibitem{Harabi07}
F.~Harabi, H.~Changuel, and A.~Gharsallah, ``Estimation of 2-d direction of
  arrival with an extended correlation matrix,'' in \emph{Positioning,
  Navigation and Communication, 2007. WPNC '07. 4th Workshop on}, March 2007,
  pp. 255--260.

\bibitem{Sujuan04}
W.~Sujuan, D.~Jiahao, P.~Shuguang, S.~Dongning, and D.~Xingguang, ``The
  estimation for 2-d direction of arrival based on higher-order cumulant of
  signals received by circle array,'' in \emph{Computational Electromagnetics
  and Its Applications, 2004. Proceedings. ICCEA 2004. 2004 3rd International
  Conference on}, Nov 2004, pp. 348--351.

\bibitem{Albagory13}
Y.~Albagory and A.~ashour, ``Music 2d-doa estimation using split vertical
  linear and circular arrays,'' \emph{Computer Network and Information Security
  (IJCNIS), International Journal of}, vol.~5, no.~8, pp. 12--18, June 2013.

\bibitem{LiBook07}
L.~Tan, \emph{Digital Signal Processing Fundamentals and Applications}.\hskip
  1em plus 0.5em minus 0.4em\relax Academic Press, 2007.

\bibitem{Brassard94}
G.~Brassard and L.~Salvail, ``Secret-key reconciliation by public
  discussion.''\hskip 1em plus 0.5em minus 0.4em\relax Springer-Verlag, 1994,
  pp. 410--423.

\end{thebibliography}
\end{document}